# Frequency Vectoralization and Frequency Birefringence

Alireza Akbarzadeh, Nima Chamanara, and Christophe Caloz

Department of Electrical Engineering, École polytechnique de Montréal, Montréal, QC, H3T 1J4 Canada

Email: alireza.akbarzadeh@polymtl.ca

**Abstract.** *In view of momentum continuity at a temporal slab, it is shown that instantaneous switching of an isotropic medium to an anisotropic medium offers the incident frequency a directional property- a counterintuitive process which is called frequency vectoralization. By expressing the dispersion diagrams before and after the temporal transition, a general analogy between spatial and temporal interfaces is given and the concept of frequency birefringence, i.e. double frequency jump, will be explained. Furthermore, it will be shown that an anisotropic temporal interface diffracts a monochromatic beam in both the spatial and spectral domains.*

Symmetry, i.e. the invariance of physical entities such as energy and momentum with temporal or spatial transformations, has been one of the most intriguing subjects of study in physics and mathematics. As discovered by Emmy Noether in 1918 [1, 2], any universal continuous symmetry is associated with a conservation law in physics. For instance, the conservation of momentum is a consequence of spatial symmetry, and the conservation of energy is a result of temporal translational symmetry.

However, regardless of the fact that most of the fundamental laws of physics are symmetric, as has been sought considerably from ancient times till now, most of the physical systems are not globally or continuously symmetric, and symmetry breaking is the cause of the majority of physical phenomena in nature. Refraction and reflection (or in general scattering) of light are two obvious examples of spatial translational symmetry breaking. Crystals, electronic and photonic band gaps are products of spontaneous breaking of continuous spatial translational symmetry into discrete spatial translational symmetry. Mass formation, superfluidity, spin waves propagation, quantum and classical Hall effects, Casimir effect, and electromagnetic radiation are several other examples appeared due to the spatial symmetry breaking [2].

Contrary to the spatial symmetry breaking, the temporal symmetry breaking and light propagation in temporally inhomogeneous media, has not been scrutinized as much. Probably Albert Einstein was the first person who considered the invariance of laws of physics with the temporal variation in his celebrated theory of special relativity, and connected space and time symmetry breaking via the Lorentz transformations. Space-time media in general are divided into two categories: 1) mechanically moving media, 2) spatiotemporally modulated media. The electrodynamics of the former category has been studied extensively and a rich literature concerning various phenomena in media at motion is available [3-5]. However, due to several undesired properties of moving media, such as difficulty to reach high speeds in practice, existence of Fizeau drag, and bi-anisotropy, they are not favoured in many practical applications.

Regarding the latter category, Morganthaler for the first time formulated the problem of light propagation in a medium with an instantaneous step-like temporally varying refractive index [6], and around a decade later his work was extended by Falsen and Whitman [7] and by Fante [8]. Then during years the optics of nonstationary media were mainly the subject of study in plasma physics [9] and nonlinear processes [10]. However, recently a renewed attention has been paid to the physics of light-matter interaction in linear space-time media. A series of paper by Agrawal and his co-authors revisited various aspects such as transmission and reflection as well as continuity conditions at a temporal boundary [11], modal analysis in temporal resonators [12], and pulse propagation in fiber optics with time-varying refractive indices [13]. Furthermore, time-reversal and light trapping in temporally modulated photonic crystals [14], light scattering and reversed nonrelativistic Doppler effect in photonic crystals with shochlike dielectric modulations [15], temporal cloaks (or history cloaks) and spatiotemporal transformation optics [16] have been studied in last two decades. Biancalana *et al*. showed how generally the electromagnetic wave propagation should be formulated in space-time media with subluminal and superluminal modulation [17]. The spectrum of plasmonic modes in space-time modulated graphene, the leaky-wave systems with spatiotemporal modulation, and non-stationary metasurfaces were considered in [18]. From quantum perspective, topics like electromagnetic quantization, squeezed and correlated quantum states, time refraction and photon acceleration, and quantum radiation in space-time media have been considered in [19].

In all the above-mentioned works, the space-time symmetry is broken such that the energy and/or momentum of light at the spatiotemporal discontinuity change in a desired manner. Likewise, in this article, we are presenting a specific type of temporal symmetry breaking, which leads the frequency (energy) of the incident light to behave differently for different angles of incidence. In other words, the introduced discontinuity makes the incident frequency (energy) distinguishes between different geometrical directions. This is a unique phenomenon that we prefer to call "frequency vecotoralization". As will be explained in details, depending on the polarization state of incident light and as a consequence of phase matching, the proposed temporal interface reshape the isotropic space of incoming frequency into two folds: 1) an isotropic space for the out-of-plane polarization ($s$ polarization), i.e. the wave vector is along the plane spanned by the magnetic field components, 2) an anisotropic elliptical shell for the in-plane polarization ($p$ polarization), i.e. the wave vector is along the plane spanned by the electric displacement field components. Based on this analysis, we present a *Snell law* analogy for frequency refraction in a birefringent manner. Then we proceed with studying how a diffractive wave may interact with the suggested interface and how the spectral components of such fields redistribute in the spectral domain.

Assume we have an initial unbounded dielectric medium with the refractive index of $n = n_1$ and without loss of generality let us assume a monochromatic plane wave with the frequency $\omega_1$ and wave vector $\vec{k}_1 = \hat{e}_x k_x + \hat{e}_z k_z$ is propagating along the $x-z$ plane. Hence, the dispersion relation for the plane wave is simply written as,

$$\omega_1^2 = \omega_x^2 + \omega_z^2 \tag{1}$$

where, $\omega_x = (c/n_1)k_x$, $\omega_z = (c/n_1)k_z$, and $c$ is the speed of light in vacuum. As shown in Figure 1, the corresponding dispersion diagram on the $\omega_x - \omega_z$ space is a circle with its radius fixed at $\omega = \omega_1$, which trivially means that along all the directions (i.e. for all the possible wave vector components) the frequency remains the same.

Now, assume by some means of external excitation we instantaneously switch the initial unbounded isotropic medium to a uniaxial medium with the optical axis parallel to the $z$ axis and refractive index tensor $\ddot{n} = diag\{n_\parallel, n_\parallel, n_\perp\}$, where $diag$ stands for the diagonal tensor. As discussed in details in [11], due to the continuity of free charge and free currents, by temporal transitions, the electric displacement and magnetic field vectors remain conserved. Since they make a right triplet with the wave vector, then the wave vector components do not change at the proposed temporal interface. In other words, as the considered interface breaks the temporal translational symmetry, the momentum of the electromagnetic waves must remain unchanged before and after the interface, while the energy alters abruptly [17]. Then, with the use of the wave vector continuity, the dispersion relation in the uniaxial state for the out-of-plane (or ordinary) polarization, i.e. the electric field being parallel to the $y$ axis, reads

$$\omega_\parallel^2 = (n_1/n_\parallel)^2 (\omega_x^2 + \omega_z^2) \tag{2}$$

which simplifies to,

$$\omega_\parallel = (n_1/n_\parallel)\omega_1 \tag{3}$$

This is the conventional Doppler effect which has been used in all the cited papers regarding the temporal discontinuities. As shown in Figure 1, the dispersion diagram in $\omega_x - \omega_z$ plane is again simply a circle with its constant radius being the frequency of propagation in all the directions. However, for the in-plane (extraordinary) polarization, i.e. the electric field being along the $x - z$ plane, the dispersion relation is

$$\omega_\perp^2 = (n_1/n_\perp)^2 \omega_x^2 + (n_1/n_\parallel)^2 \omega_z^2 \tag{4}$$

The related dispersion shell is illustrated in Figure 1, where the length of traced vectors connecting the origin to the shell are the frequency values after the transition. As seen in equation (4) and depicted in Figure 1, the frequency jump of the in-plane polarization after the transition is a function of the wave vector, which means that the refracted frequency accepts values between the two limiting cases of $(n_1/n_\perp)\omega_1$ and $(n_1/n_\parallel)\omega_1$. In other words, through the proposed temporal transition, the refracted frequency of the in-plane polarization distinguishes between various initial propagation directions and, similar to a vector field, behaves differently for different directions- we would like to call the temporal process which causes this counterintuitive frequency change, "frequency vectoralization".

A constructive demonstration of the introduced frequency vectoralization is an analogy with the spatial transition counterpart. When a light ray faces an anisotropic spatial symmetry breaking, it splits into two rays with two different phase velocities- a phenomenon well-known as birefringence or double refraction. One conventionally can determine the direction of the refracted rays through the continuity of tangential components of incident and refracted wave vectors on the isofrequency dispersion shells in the $\vec{k}$ space. This is essentially an immediate consequence of energy continuity and nonconservation of momentum at the spatial interfaces. As mentioned earlier, the counterpart of spatial symmetry is the temporal interface, at which the momentum is conserved and energy alters respectively. As a matter of fact, the wave vector remains unchanged, while the frequency changes correspondingly. Hence, in view of birefringence at the spatial interfaces, the frequency of the light ray incident at an anisotropic temporal interface splits into two branches, which we call it frequency birefringence or frequency double refraction. In order to determine the values of the refracted frequencies at the considered temporal interface, one should essentially consider the iso-wave-vector contours over the frequency space and apply the wave vector continuity. Accordingly the Snell's law for the frequency birefringence can be written as

$$n_1 \omega_1 = n_\parallel \omega_\parallel \tag{5}$$

for the out-of-plane polarization, and

$$n_1 \omega_1 = \left[ n_\parallel^2 \sin(\theta_i)^2 + n_\perp^2 \cos(\theta_i)^2 \right]^{-0.5} n_\parallel n_\perp \omega_\perp \tag{6}$$

for the in-plane polarization, where $\theta_i$ is the angle of propagation. The angle-dependent frequency and frequency birefringence are schematically illustrated in Figure 2. As seen in Figure 2(a), two in-plane polarized light rays with the frequency of $\omega_1$ and propagation angles of $\theta_{1a}$ and $\theta_{1b}$ are propagating in an initially isotropic medium. Then by temporally transforming the medium into a uniaxial medium, the two rays continue their trajectories without any change in their angles, while their frequencies jump into different values, which are obtained from equation (6). Shown in Figure 2(b) is the frequency birefringence of an unpolarised light ray incident onto a uniaxial temporal interface. At the instant of transition, the in-plane and out-of-plane components continue their trajectories at the same direction as the incident light, but with frequencies $\omega_\perp$ and $\omega_\parallel$, respectively.

In view of the angle-dependent frequency splitting and frequency birefringence, it is noteworthy to study the temporal diffraction of spatially diffractive light beams. Consider a monochromatic diffractive beam with frequency $\omega_1$ is propagating along $z$ axis in an initially isotropic medium. With the use angular spectrum analysis, the electric field of the beam can be essentially written as a superposition of infinite plane waves,

$$\vec{E}(x,z;t) = \int_{-\infty}^{+\infty} dk_x \hat{E}(k_x) e^{-ik_x x - ik_z z} e^{+in_1 \omega_1 t} \tag{7}$$

where $\hat{E}$ can be decomposed into in-plane and out-of-plane components as, $\hat{E} = \hat{E}_{\parallel} + \hat{E}_{\perp}$. Then after the uniaxial temporal transition, similar to any other temporal transition [17], a forward and a backward mode for each of the polarizations are produced and, hence the spectral decomposition of the electric field reads,

$$\vec{E}(x,z;t) = \int_{-\infty}^{+\infty} dk_x \hat{E}_{\parallel}(k_x) e^{-ik_x x - ik_z z} \left( e^{+i\omega_{\parallel} t} + e^{-i\omega_{\parallel} t} \right)$$
$$+ \int_{-\infty}^{+\infty} dk_x \hat{E}_{\perp}(k_x) e^{-ik_x x - ik_z z} \left( e^{+i\omega_{\perp}(k_x,k_z)t} + e^{-i\omega_{\perp}(k_x,k_z)t} \right)$$
(8)

where the first and the third terms are regarding the forward modes, and the second and fourth terms are concerning the backward modes. As seen in equation (8), the incident beam splits into two beams at the considered transition. The out-of-plane part, i.e. the first integral on the right hand side of equation (8), maintains its initial envelope with a single frequency transition from $\omega_1$ to $(n_1/n_{\parallel})\omega_1$. However, for the out-of-plane polarization, the frequency transition for the respective spectral components is inhomogeneous, which leads to a polychromatic new beam which may have an envelope different from that of the initial beam. In order to illustrate the monochromatic to polychromatic transition, a cylindrical plane wave with uniform distribution of the spectral components and a single frequency, i.e. blue color, is considered (see Figure 3). As seen in Figure 3(a), the out-of-plane part of the incident field undergoes an abrupt change in its frequency, i.e. blue to red transition. However, as shown in Figure 3(b), for the in-plane part the initial frequency diffracts into a continuous range of frequencies over the interval between $n_1\omega_1/n_{\perp}$ (green) and $n_1\omega_1/n_{\parallel}$ (red). It should be noted for the limiting case of propagating along the optical axis $\vec{k} = \hat{e}_z k_z$ the in-plane and out-of-plane frequency transitions overlap, which can be inferred from Figure 1(b), where the green and pink frequency shells are tangent to each other at $\omega_x = 0$. This angular spectrum analysis reveals that the temporal interface not only refract the incident frequency, but also it causes a spectrum diffraction- a phenomenon which can be explored for useful telecommunication applications.

In conclusion, we studied the frequency refraction and diffraction at a temporal interface between an isotropic dielectric medium and a uniaxial dielectric medium. We showed that the frequency of the in-plane polarization refracts according to the direction of propagation, which can be of great use in many applications. Through a general analogy between spatial and temporal symmetry breaking, we offered a Snell law for frequency birefringence. Based on the provided concepts, we showed how a temporal interface may diffract a beam both in spatial and spectral domains.

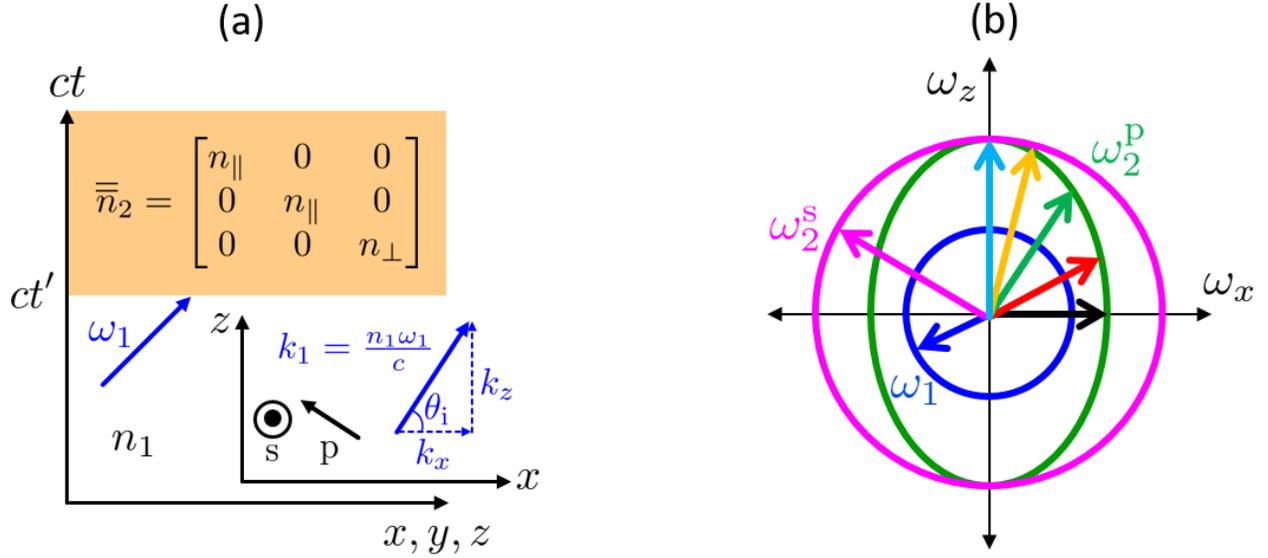

FIG. 1. (a) The temporal transition at $ct = ct'$ from an isotropic dielectric medium with refractive index $n = n_1$ to a uniaxial medium with refractive tensor $\bar{\bar{n}} = \bar{\bar{n}}_2$. The incident plane wave has the frequency $\omega = \omega_1$ and angle of propagation $\theta_i$ with in-plane ($p$) and/or out-of-plane ($s$) polarization. (b) The dispersion diagram of the isotropic and anisotropic before and after the temporal transition in $\omega_x - \omega_z$ space. The blue circle is the dispersion shell of the initial isotropic medium with radius $\omega = \omega_1$. The pink circle and the green ellipse are the dispersion shells for the out-of-plane polarization and in-plane polarization after the transition, respectively. The vector connecting the origin to the pink circle $\omega_{2s}$ (the out-of-plane frequency) is of the same length, while the length of the vector connecting the origin and the green circle is changing according to the angle of propagation.

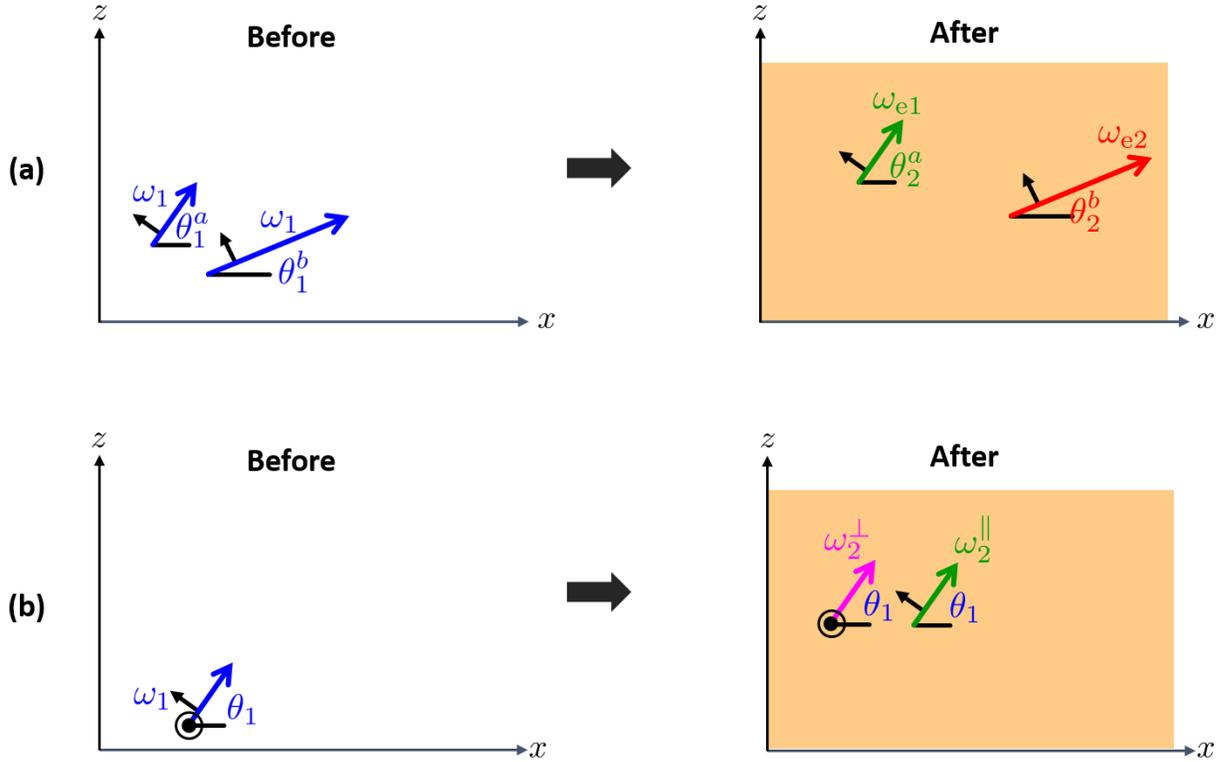

FIG. 2. (a) The angle-dependent frequency change at the temporal interface. Before the transition the two rays have the same frequency and are in-plane polarized (black arrow). After the transition, the two ray continue exactly the same trajectory with the same polarization, while the change in their frequencies are different according to their propagating angles. (b) The frequency birefringence at the uniaxial temporal interface: an unpolarised light with frequency $\omega = \omega_1$ is propagating at the angle $\theta = \theta_1$. After the transition, the incident ray split into an in-plane (green ray) and an out-of-plane (pink ray) polarized ray. The polarized rays have identical trajectories, while their frequencies are different.

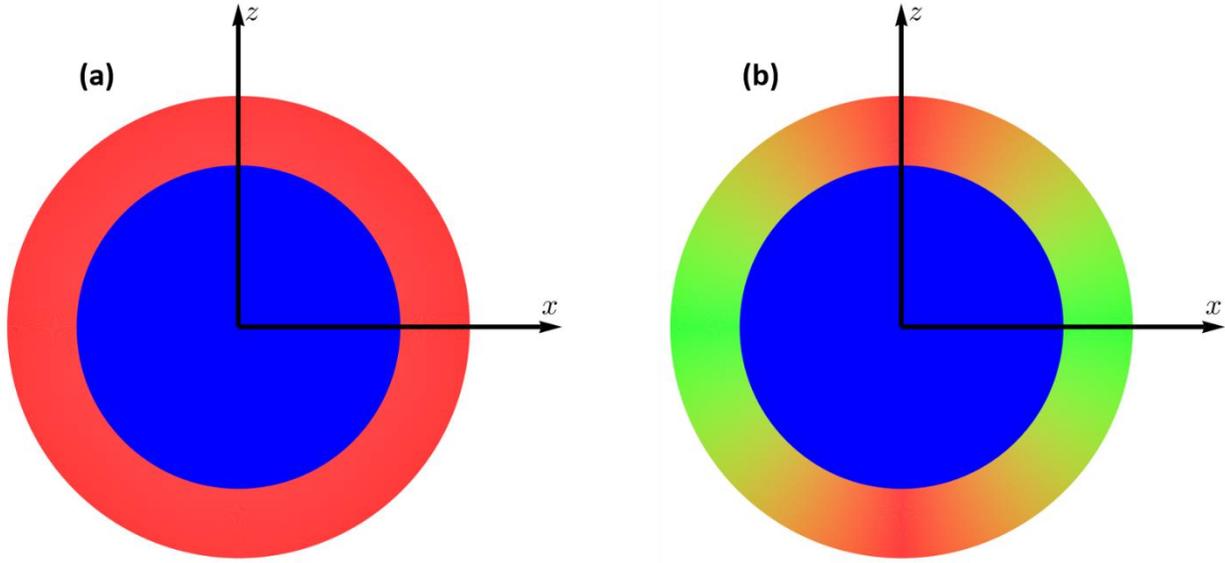

FIG. 3. The spectrum of a uniformly propagating cylindrical wave with (a) out-of-plane, and (b) in-plane polarization before and after the uniaxial temporal transition. The out-of-plane polarization just jumps from the blue frequency $\omega_1 = \omega_{blue}$ to the red one. But the in-plane polarization jumps from blue to all the frequencies between the limiting cases of the green $\omega_{green} = (n_1/n_\perp)\omega_{blue}$ and red $\omega_{red} = (n_1/n_\parallel)\omega_{blue}$.